\documentclass[10pt,a4paper]{article}

\usepackage{a4} 
\usepackage{ucs}
\usepackage[utf8x]{inputenc}
\usepackage{parskip} 
\usepackage{amsmath} 
\usepackage{amssymb} 
\usepackage{graphicx} 
\usepackage{cite} 
\usepackage{mathpazo} 
\usepackage[scaled=.92]{helvet} 
\usepackage{color} 
\usepackage[colorlinks=true]{hyperref}

\definecolor{dblue}{rgb}{0,0,0.5}
\definecolor{dgreen}{rgb}{0,0.5,0}
\definecolor{dred}{rgb}{0.5,0,0}
\hypersetup{linkcolor=dgreen,pagecolor=dgreen,urlcolor=dblue,citecolor=dred}

\begin{document}


\title{A measure for the chirality of triangles}

\author{ Haifeng Ma and Thomas Greber}
\maketitle
\ {Physik Institut Universit\"at Z\"urich, Winterthurerstrasse 190, CH-8057 Z\"urich, Switzerland}

\ {greber@physik.uzh.ch}
\begin{abstract}
{A measure for the description of the chirality of triangles is introduced.
The measure $\chi_\vartriangle$ is zero for triangles with at least one mirror axe, i.e.  equilateral or isosceles triangles, and positive or negative for scalane, i.e. left or right handed triangles, respectively.}
\end{abstract}

\begin{figure}[h] 
\centering
\includegraphics[width=7 cm]{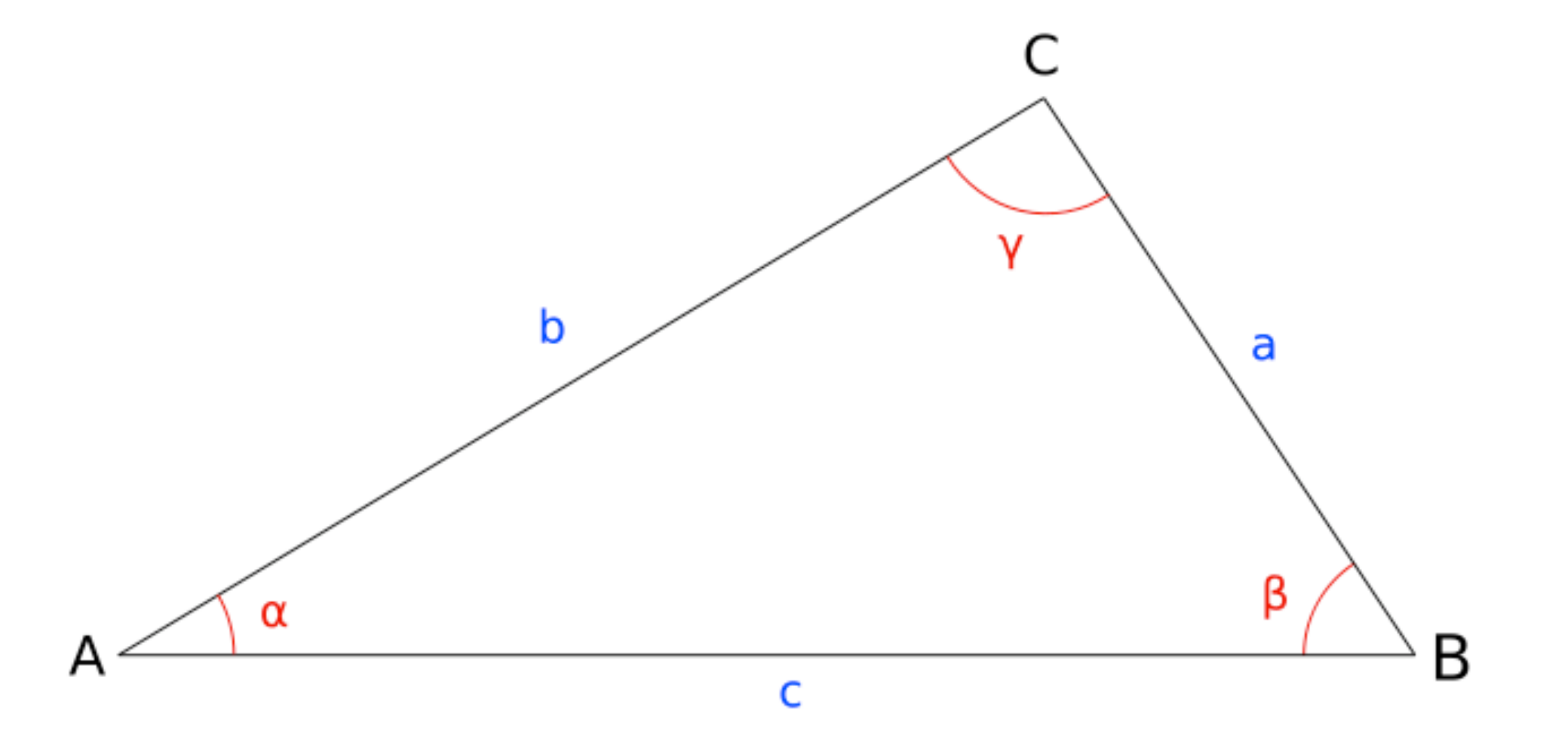}

\caption{Triangle and the definition of the standard notation (From Ref. \cite{triangle}).}
\label{F1}
\end{figure}

A triangle is a polygon with three vertices $A,B,C$ connected by three sides (line segments) $a,b,c$. Figure \ref{F1} shows the definition of the symbols \cite{triangle}. Note that $A,B,C$ are labeled in a counter clock wise fashion, if we look from the top.
It is a two dimensional object since 3 points define a two dimensional plane.

When we analyze the symmetry of triangles we find {\emph{equilateral}} triangles with $a=b=c$, which have $C_{3v}$ symmetry, i. e. three fold rotational symmetry and 3 mirror axes.
The {\emph{ isosceles}} triangles have two sides with equal length, and $C_{1v}$ symmetry, i.e. one fold rotational symmetry and 1 mirror axe.
All other triangles, where $a \neq b\neq c \neq a$ are {\emph{scalane}} and have no mirror symmetry.
Since the concept of chirality and mirror symmetry may also be applied to two dimensional objects, scalane triangles are considered to be chiral. In two dimensions, the role of a mirror plane in three dimensional objects is taken by a mirror axis. 

\subsection*{A measure for the chirality of a triangle $\chi_\vartriangle$ }

There are different possible sets for the definition of the chirality of a geometric object \cite{Petitjean:2003}.
Here we define a measure for the chirality of a triangle $\chi_\vartriangle$ as :

\begin{equation}
\chi_\vartriangle=  \frac{(a-b)\cdot(b-c)\cdot(c-a)}{(a+b+c)^3}
\label{E0}
\end {equation}

where $a,b,c$ are the sides of the triangle, with the sense of rotation as defined in Figure \ref{F1}.
For example, the 3-tuple $(a,b,c)=(9,10,11)$ has a $\chi_\vartriangle$ of $7.4 \cdot 10^{-5}$.
$\chi_\vartriangle$ is a dimensionless number and a bound quantity $-\chi_{max} \leq \chi_\vartriangle \leq+\chi_{max}$, with  $\chi_{max}\approx \frac{1}{83}$.
For triangle $\chi_\vartriangle$ gets maximal (and minimal) when one side of the triangle is half of  the circumference ($a+b+c$). 
For example, the tuple (10, 2, 8) represents such a triangle with $a=b+c$, which is with $\chi_\vartriangle \approx 1.2 \cdot 10^{-2}$ close to $\chi_{max}$.

\begin{figure}[h!] 
\centering
\includegraphics[width=7 cm]{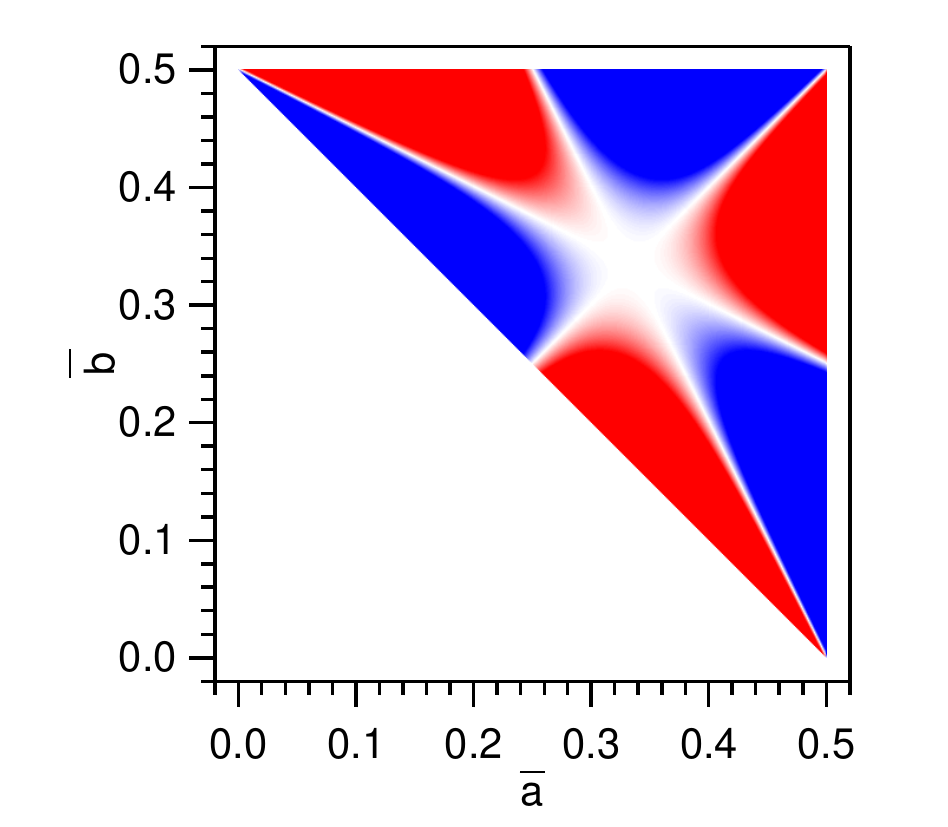}
\caption{Plot of the phase space of $\chi_\vartriangle (a,b,c)$ for normalised triangle sides ($\bar{a},\bar{b},\bar{c}$), with $\bar{c}=1-\bar{a}-\bar{b}$. Blue are positive $\chi_\vartriangle$, e.g. for $a\geq c \geq b$.  Red are negative $\chi_\vartriangle$ and indicate the opposite chirality. Note the white nodal lines where $\chi_\vartriangle =0$. Near the equilateral triangle with $\bar{a}=\bar{b}=\bar{c}=\frac{1}{3} $, $|\chi_\vartriangle|$ is small and approaches zero.}
\label{F2}
\end{figure}

In Figure \ref{F2} the phase space of  $\chi_\vartriangle (a,b,c)$ is shown as a function of ($\bar{a},\bar{b}$), where $\bar{a}$, $\bar{b}$ and $\bar{c}$ are $a$, $b$ and $c$ normalized with the circumference ($a+b+c$), and where $\bar{c}=1-\bar{a}-\bar{b}$.
The 3 blue and 3 red areas, are separated by  3 nodal lines of isosceles triangles. 
The 6 segments represent the $3!$ possible permutations of the elements in a 3-tuple, ($a,b,c$), ($c,a,b$), ($b,c,a$), ($b,a,c$), ($c,b,a$) and ($a,c,b$).
Note, for cyclic permutations we find  $\chi_\vartriangle (a,b,c)=\chi_\vartriangle (c,a,b)=\chi_\vartriangle (b,c,a)$ and permutations of pairs lead to a change of sign of the chirality
$\chi_\vartriangle (a,b,c)=-\chi_\vartriangle (b,a,c)$, etc.

If triangles are classified according to Eqn. \ref{E0}, and if we assign the largest element to $a$, the phase space of $\chi_\vartriangle$ is reduced by a factor of 3, i.e. to one blue and one red segment.
If we order the labels of the sides to e.g. ($a\geq b \geq c$) we loose the information on the absolute chirality and can display all 3-tuples in one, for this case red segment in Figure \ref{F2}.

\subsection*{Generalisation of Eqn.\ref{E0} to any 3-tuple}
The definition of $\chi_\vartriangle$ (Eqn.\ref{E0}) also applies to the chiral classification of any three scalar quantities (3-tuples ($a,b,c$) with $a>0$, $b>0$, $c>0$). 
Though, for 3-tuples that do not fulfil the triangle inequality as it is e.g. ($a,b,c$)=(1, 2, 8), $\chi_\vartriangle$ gets 0.03, i.e.  it can get larger than $\chi_{max}$.

\subsection*{Which $\chi_\vartriangle$ indicate a significant chirality}

In an experiment ($a,b,c$) can be measured up to an error ($\Delta a, \Delta b, \Delta c$).
This imposes an error in $\chi_\vartriangle$ that has to be known if it shall be decided whether  $\chi_\vartriangle$ is significantly different from zero, i.e. whether a tuple ($a,b,c$) may be considered to be chiral.
It has therefore to be studied the influence of an error ($\Delta a, \Delta b, \Delta c$) on $\chi_\vartriangle$.

\begin{figure}[h!] 
\centering
\includegraphics[width=13 cm]{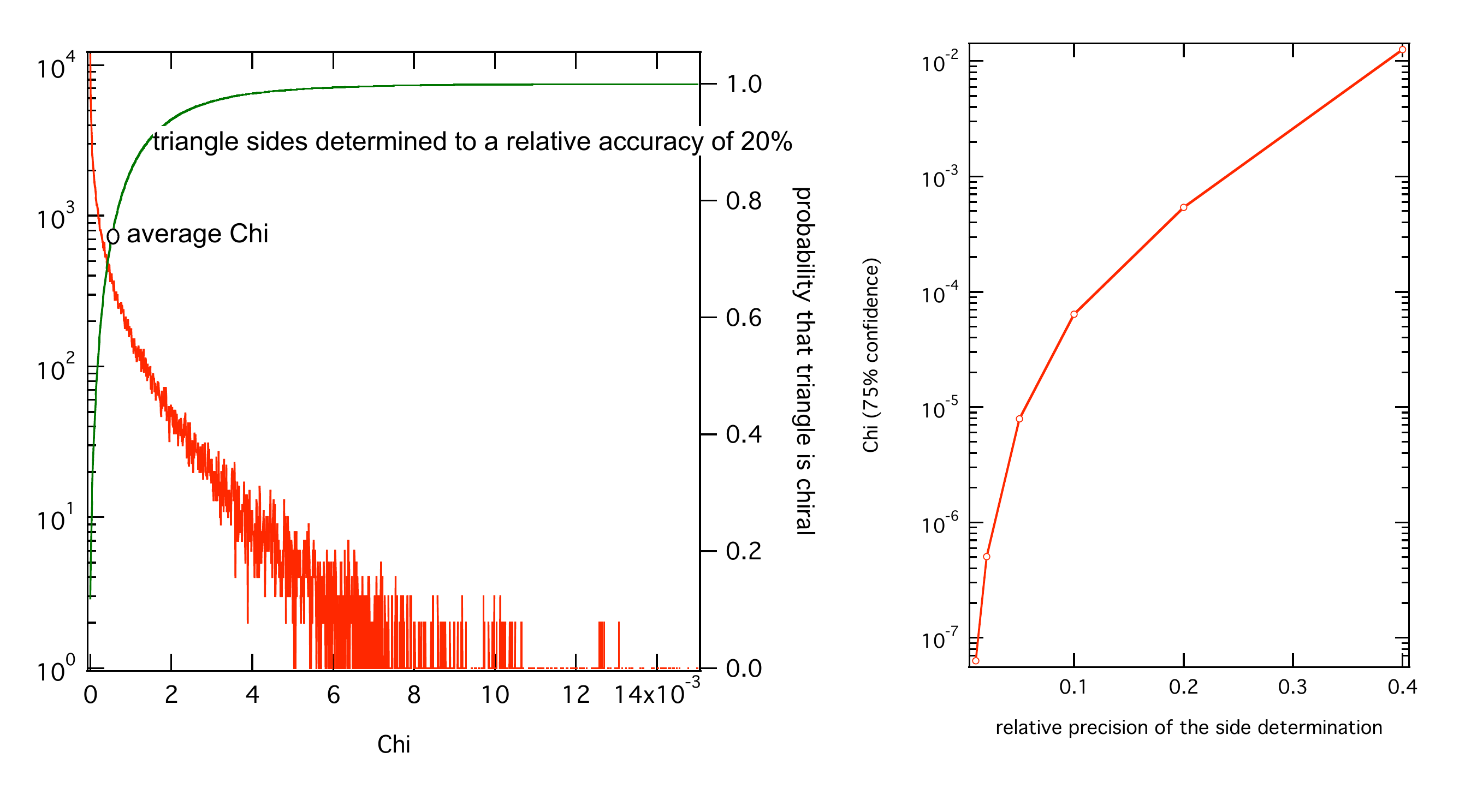}
\caption{Plot of the distribution of $|\chi_\vartriangle|$ for  equilateral triangles with sides equal to 5+gnoise(1) (left panel) ($10^5$ samples). In the right panel the average of $|\chi_\vartriangle|$ is shown as a function of the relative accuracy of the side determination $\frac{\Delta a}{a}$.  The errors are simulated with a Gaussian distribution (gnoise(y)), where y is the standard deviation of the distribution. The integral of the histogram (green line), right axis, shows that about 75\% of the $|\chi_\vartriangle|$ lie below the average of the distribution, which is $5.5\cdot10^{-4}$ for the case of 20\% error. 
The right panel shows this 75\% confidence values for $\chi_\vartriangle$'s with different errors (relative precision of the side determination) (see Table \ref{T1}).}
\label{F3}
\end{figure}

In a scenario for equilateral triangles $a=b=c$ with relative random errors $\frac{\Delta a}{a}=\frac{\Delta b}{b}=\frac{\Delta c}{c}$ = 0.2 we get for $|\chi_\vartriangle|$  the histogram shown in Figure \ref{F3}.
75\% of the samples have a $\chi_\vartriangle$ below the average $\bar{\chi}_\vartriangle$.
For this particular simulation with $10^5$ samples, also $\chi$-values $>\chi_{max}$ occur. 
This is related to the fact that certain 3-tuples can not be considered as sides of a triangle, since if sorted like $a > c \geq b$ we find $a-b-c \geq 0$, i.e. these  samples  (about 0.5\% for 20\% errors) do not fulfil the triangle inequality.
On the right panel in Figure \ref{F3} and in Table \ref{T1} the 75\% confidence values for different errors (relative precision of the side determination) are shown. 
This means for a triangle with $\chi_\vartriangle \geq6\cdot10^{-5}$, whose sides are measured with 10\% accuracy that it is with a probability of 75\% chiral, and with a probability of 25\% achiral.

\begin{table}
\centering
\begin{tabular}{r|r}
\hline Error & $\chi_\vartriangle$ \\ 
\hline 0.01 & $6.27 \cdot 10^{-8}$ \\ 
\hline 0.025 & $9.75 \cdot 10^{-7}$ \\ 
\hline 0.05 & $7.83 \cdot 10^{-6}$ \\ 
\hline 0.075 & $2.67 \cdot 10^{-5}$ \\ 
\hline 0.10 & $6.40 \cdot 10^{-5}$ \\ 
\hline 0.15 & $2.22 \cdot 10^{-4}$ \\ 
\hline 0.20 & $5.51 \cdot 10^{-4}$ \\ 
\hline 0.25 & $1.12 \cdot 10^{-3}$ \\ 
\hline 0.30 & $2.10 \cdot 10^{-3}$ \\ 
\hline 0.35 & $3.76 \cdot 10^{-3}$ \\ 
\hline 0.40 & $7.11 \cdot 10^{-3}$ \\ 
\end{tabular}

\caption{75\% confidence values for $\chi_\vartriangle$ to be chiral for different relative errors $\frac{\Delta a}{a}=\frac{\Delta b}{b}=\frac{\Delta c}{c}$. For details see text.}
\label{T1}
\end{table}\

\end{document}